\documentclass[twocolumn,aps,superscriptaddress,showpacs]{revtex4}
\usepackage{amssymb}
\usepackage{amsmath}
\usepackage{graphicx}
\usepackage[normalem]{ulem}
\usepackage[dvips]{color}

\begin{document}

 \title{GW170817 implications on the frequency and damping time of f-mode oscillations of neutron stars}
\author{De-Hua Wen}
\affiliation{School of Physics and Optoelectronics, South China University of Technology, Guangzhou 510641, P.R. China}
\affiliation{Department of physics and Astronomy, Texas A$\&$M University-Commerce, Commerce, TX 75429, USA}
\author{Bao-An Li\footnote{Corresponding author: Bao-An.Li@Tamuc.edu}}
\affiliation{Department of physics and Astronomy, Texas A$\&$M University-Commerce, Commerce, TX 75429, USA}
\author{Hou-Yuan Chen}
\affiliation{School of Physics and Optoelectronics, South China University of
Technology, Guangzhou 510641, P.R. China}
\author{Nai-Bo Zhang}
\affiliation{Shandong Provincial Key Laboratory of Optical Astronomy and Solar-Terrestrial Environment, Institute of Space Science, Shandong University, Weihai 264209, Shandong, China}

\begin{abstract}
Within a minimum model for neutron stars consisting of nucleons, electrons and muons at $\beta$-equilibrium using about a dozen Equation of States (EOSs) from microscopic nuclear many-body theories and 40,000 EOSs randomly generated using an explicitly isospin-dependent parametric EOS model for high-density neutron-rich nucleonic matter within its currently known uncertainty range, we study correlations among the f-mode frequency, its damping time and the tidal deformability as well as the compactness of neutron stars.  Except for quark stars, both the f-mode frequency and damping time of canonical neutron stars are found to scale with the tidal deformability independent of the EOSs used.
Applying the constraint on the tidal deformability of canonical neutron stars $\Lambda_{1.4}=190^{+390}_{-120}$ extracted by the LIGO+VIRGO Collaborations from their improved analyses of the GW170817 event, the f-mode frequency and its damping time of canonical neutron stars are limited to 1.67 kHz - 2.18 kHz and 0.155 s - 0.255 s, respectively, providing a useful guidance for the ongoing search for gravitational waves from the f-mode oscillations of isolated neutron stars. Moreover, assuming either or both the f-mode frequency and its damping time will be measured precisely in future observations with advanced gravitational wave detectors, we discuss how information about the mass and/or radius as well as the still rather elusive nuclear symmetry energies at supra-saturation densities may be extracted.
\end{abstract}


\maketitle

\section{Introduction}
If a neutron star is disturbed by an external or internal event, it may then oscillate non-radially and thus emit gravitational waves (GWs). The latter are generally expected to provide useful information about the structure and the underlying EOS of neutron stars. For a recent review on how/what information about the EOS of dense neutron-rich matter can be extracted from studying the peak frequency of post-merger GW spectrum, frequencies of both quasi-equilibrium and resonant tides in merging neutron star binaries as well as frequencies of various oscillating modes of isolated neutron stars, we refer the reader to ref. \cite{f-custipen}. Hopefully, the predicted features,  see, e.g., those summarized in refs. \cite{Mot18,Blazquez2014}, of GWs from quasi-normal modes will be verified in the near future with advanced GW detectors, such as the  Einstein Telescope \cite{Pitkin2011,sensitivities}.

The theoretical formalism for describing the GWs from the quasi-normal oscillations of neutron stars has been well established, see, e.g., refs. \cite{Lindblom83,Detweiler85}.
The quasi-normal modes can be classified into the polar and axial modes: the polar modes correspond to zonal compressions while the axial modes induce differential rotation in the fluid \cite{Andersson1996}.  In neutron stars, the polar modes can be coupled to fluid oscillations (the fundamental f-modes, pressure p-modes, and a branch of space-time modes: the polar w-modes) \cite{Kokkotas1999,Lau2010} while the axial modes are purely the space-time modes of oscillations (the axial w-modes) \cite{Chandrasekhar1991,Kokkotas1992}. Here we focus on the f-mode as its relatively low frequency (1$\sim$ 3 kHz) makes it relatively easier to be observed than other modes \cite{Lau2010}. For example, for a neutron star located at 10 kpc from us, Kokkotas et al. \cite{Kokkotas2001} estimated that the energy required in the f-mode in order to be detected with a signal-to-noise ratio of 10 by the advanced LIGO detector is
$8.7\times 10^{-7} M_{\odot}$. It is far smaller than that for the p-mode and w-mode.

There are several different ways to stimulate the pulsation modes \cite{Kokkotas2001}: (a) A supernova explosion. An optimistic estimate for the energy radiated as the GW from supernovae can be up to
about $10^{-2}$ M$_{\odot}$.
(b) A starquake associated with a pulsar glitch. The typical
energy released in this process is estimated to be about $10^{-9}-10^{-7} ~M_{\odot}$ \cite{Mock1998}.
(c) A binary neutron star merger remnant. Before the remnant collapses to a black hole, it is expected that several oscillation modes can be excited \cite{Baumgarte1996}. Moreover, the oscillation modes of individual neutron stars
can also be excited by the tidal fields before they merger \cite{Kokkotas1995}.
(d) A prominent phase transition, which can lead
to a mini-collapse in neutron stars and thus results in a sudden softening of the EOS. An optimistic estimate for the energy radiated as GWs could be up to  $10^{-2}$ M$_{\odot}$\cite{Kokkotas2001,Lin2011}.

While the above expectations/estimates are based on solid theoretical studies, essentially none of them have been observationally confirmed yet. Thus, relating them with existing observations is very useful. The first detection of a binary neutron star merger event GW170817 has opened a new window for understanding properties and the EOS of neutron stars. Indeed, many interesting physics, see, e.g., refs. \cite{Margalit2017,Rezzolla2018,Ruiz2018,LIGO2018,Annala2018,Most2018,De2018,Chakravarti2019}, has been extracted from the historical event GW170817. One key quantity extracted from the GW170817 is
the tidal deformation $\Lambda_{1.4}=190^{+390}_{-120}$ \cite{LIGO2018} for canonical neutron stars of mass 1.4M$_{\odot}$.
The f-mode induced by the tidal force during mergers of neutron star binaries has also been investigated in recent years, see, e.g. refs. \cite{Gold2012,Chirenti2017,Papenfort2018}.
It would thus be interesting to study if/how the extracted tidal deformability may
help constrain any features of the f-mode oscillations of neutron stars. We find that the f-mode frequency and damping time for canonical neutron star are limited to 1.67 kHz - 2.18 kHz and 0.155 s - 0.255 s, respectively. As refined measurements of the tidal deformability for more merger events are expected, tighter constraints on the f-mode will come. In turn, future measurements of the f-mode frequency and damping time themselves will help crosscheck results from other approaches in the era of multi-messenger astronomy.

It is easy to understand that GWs emitted from the quasi-normal modes carry useful information about global properties and the internal structure as well as the EOS of neutron stars. But how can we decipher this information from the detected GWs? Normally, one investigates features, such as the frequency and damping time, of quasi-normal modes through numerically solving differential equations governing the perturbed metric with model EOSs of neutron star matter. If we call this route as the ``direct" way that is basically straightforward given an EOS, the ``inverse" approach (i.e., using the observed f-mode features to infer properties and the underlying EOS of neutron stars) is not easy because the related differential equations are too complicated to be numerically inverted. Besides the standard Bayesian inference, a practically useful way is to make use of the well established universal relations (independent of the EOS of neutron star matter) revealed in studying the frequency and damping time of the quasi-normal modes and their dependences on the compactness of neutron stars, see, e.g., refs. \cite{Blazquez2014,Andersson1996,Lau2010,Brazil}. Obviously, if the universal relations are correct, one can use observed GW features to determine the scaling parameters, such as the stellar mass, radius and moment of inertia. Can we further determine the underlying EOS? The answer is yes since the global properties of neutron stars depend strongly on the EOS. We shall explore if/how this may be done after verifying some of the well known universal relations.

The paper is organized as follows. In Sec. \ref{eos1}, the parametric EOS of dense neutron-rich nucleonic matter is outlined.  In Sec. \ref{fmode}, we first study the correlations between the f-mode frequency (damping time) and the tidal deformation. Constraints on the f-mode frequency and its damping time by the tidal deformation of canonical neutron stars extracted from the GW170817 event are then presented. In Sec. \ref{uni}, a brief review on the universal relations of the f-mode oscillations is given. In Sec. \ref{apl}, we discuss how information about global properties of neutron stars and the high-density behavior of nuclear symmetry energy can be extracted assuming either or both the f-mode frequency and damping time are measured accurately. A summary of the main points is given at the end.

\section{An explicitly isospin-dependent EOS for dense neutron-rich nucleonic matter}\label{eos1}
For ease of our following discussions, we briefly outline here how we construct the EOS within a minimum model for neutron stars consisting of neutrons, protons, electrons and muons at $\beta$-equilibrium.
More details can be found in ref. \cite{Zhang2018}. We use the NV EOS \cite{Negele73} for the inner crust and the BPS EOS  \cite{Baym71} for the outer crust. The transition point to the liquid core is found by studying where/when the
incompressibility of uniform neutron star matter becomes imaginary, see detailed discussions in ref. \cite{Zhang2018}. For the EOS of the core, there are many predictions based on various nuclear many-body theories using different interactions. The predicted EOSs from different theories often diverge especially at supra-saturation densities. Thus, to minimize model dependence in preparing the EOS of neutron star matter while be flexible and inclusive enough to cover all EOSs allowed by all known constraints, here we adopt the rather general parametric EOS model for neutron-rich nucleonic matter in the core \cite{Zhang2018}.
The total pressure as a function of energy density for the charge neutral $npe\mu$ matter at $\beta$-equilibrium is calculated self-consistently. As a basic input,
the nucleon specific energy $E(\rho,\delta)$ of neutron-rich matter with isospin asymmetry $\delta=(\rho_n-\rho_p)/\rho$ can be well approximated by the  empirical parabolic law as \cite{Bombaci1991,Li2008}:
\begin{equation} \label{EQ7}
E(\rho,\delta) =E_{0}(\rho)+E_{\textrm{sym}}(\rho)\cdot \delta^2+O(\delta^{4}),
\end{equation}
where $E_{0}(\rho)$ and $E_{\textrm{sym}}(\rho)$  are  the energy in symmetric nuclear matter and  the symmetry energy of asymmetric nuclear matter, respectively.
They can be conveniently parameterized as
\begin{equation} \label{EQ8}
E_{0}(\rho) =E_{0}(\rho_{0})+\frac{K_{0}}{2}(\frac{\rho-\rho_{0}}{3\rho_{0}})^{2}+\frac{J_{0}}{6}(\frac{\rho-\rho_{0}}{3\rho_{0}})^{3},
\end{equation}
\begin{eqnarray} \label{EQ9}
E_{\textrm{sym}}(\rho) &=&E_{\textrm{sym}}(\rho_{0})+L(\frac{\rho-\rho_{0}}{3\rho_{0}})+\frac{K_{\textrm{sym}}}{2}(\frac{\rho-\rho_{0}}{3\rho_{0}})^{2}\nonumber\\
&+&\frac{J_{\textrm{sym}}}{6}(\frac{\rho-\rho_{0}}{3\rho_{0}})^{3}.
\end{eqnarray}
According to the existing  knowledge on the parameters near the saturation density of nuclear matter, the most probable values of them are as follows: $K_{0}=230\pm20$ MeV, $E_{\textrm{sym}}(\rho_{0})=31.7\pm3.2$ MeV, $L=58.7\pm28.1$ MeV, and $-300\leq J_{0} \leq 400$ MeV, $-400\leq K_{\textrm{sym}}\leq 100$ MeV, $-200\leq J_{\textrm{sym}}\leq 800$ MeV, see, e.g., refs. \cite{Shlomo06,Piekarewicz10,Li13,Zhang17,Oertel17,Li17}.
The first three parameters $K_{0}, E_{\textrm{sym}}(\rho_{0})$, and $  L$ have already been constrained in their respective narrow ranges, while the last three parameters
$ J_{0}, K_{\textrm{sym}}$, and $ J_{\textrm{sym}}$ still have very large uncertainties. It is worth emphasizing that the above expressions have the following dual meanings: they are Taylor expansions near the saturation density for systems with low isospin asymmetries. For very neutron-rich systems especially at supra-saturation densities, they are simply parameterizations. Thus, the high-density parameters $ J_{0}, K_{\textrm{sym}}$, and $J_{\textrm{sym}}$ are no longer Taylor expansion coefficients, but free parameters to be determined by observations of neutron stars and/or high-energy heavy-ion reaction experiments especially with radioactive beams \cite{Zhang2018}. However, the uncertainty ranges of these parameters cited above mostly based on nuclear theory predictions provide a useful reference (the ranges of prior probability distribution functions of these parameters in Bayesian analyses) for their eventual inference from the experimental/observational data. By varying the EOS parameters, we can generate efficiently large numbers of EOSs for neutron star matter. These EOSs are first screened against the well accepted constraints, such as, the causality condition and the ability to support neutron stars with masses at least as high as 2.01 M$_{\odot}$, etc.

For comparisons, we also use 11 EOSs from predictions of microscopic nuclear many-body theories for normal or hybrid neutron stars (marked as 11 microscopic EOSs in the following text and figures) and 2 EOSs from the MIT bag model for quark stars. The 11 microscopic EOSs  include:
ALF2 of Alford et al. \cite{ALF2} for hybrid (nuclear + quark matter) stars,  APR3 and APR4 of Akmal and Pandharipande \cite{AP34}, ENG of Engvik et al. \cite{ENG}, MPA1 of Muther, Prakash and Ainsworth \cite{MPA1}, SLy of Douchin and Haensel \cite{SLy}, WWF1 and WWF2 of Wiringa, Fiks and Fabrocini \cite{WFF12}, the QMFL40, QMFL60 and QMFL80 within the Quark Mean Field model with L=40, 60 and 80 MeV, respectively, from the recent work of Zhu et al. \cite{QMF}. The two EOSs of quark stars are from the MIT bag model with
the pressure $p$ as a function of energy density $\epsilon$ given by $p(\epsilon) =1/3(\epsilon-4B)$ with the bag constant B=30 $\textrm{MeV}/\textrm{fm}^3$ and 57 $\textrm{MeV}/\textrm{fm}^3$, respectively \cite{Chodos1974,Glendenning1997}. We notice that the bag constant used here might be smaller than the values used by others in various studies, see, e.g., ref. \cite{Ors}. Nevertheless, as we shall show our qualitative conclusions are independent of the MIT bag model.

\section{Constraints on the f-mode frequency and damping time by the tidal deformation of neutron stars}\label{fmode}
Among the properties of neutron stars extracted from analyzing the GW170817 event, the tidal deformation is the most important one carrying information about the EOS of neutron star matter. The dimensionless tidal deformability  $\Lambda$ is defined using the second Love number $k_{2}$, stellar mass $M$ and radius $R$ as
\begin{equation} \label{EQ10}
\Lambda=\frac{2}{3}k_{2}\cdot(\frac{R}{M})^{5}.
\end{equation}
The $k_{2}$ depends on not only the mass $M$ and radius $R$ but also the interior structure of neutron stars. Numerically,
it is determined through a very complicated differential equation \cite{Hinderer2010} coupled to the Tolman-Oppenheimer-Volkov (TOV) equation \cite{TOV} for a given EOS.
The code we used to calculate the $k_{2}$ is the same one used previously in refs. \cite{Fattoyev2013,Fattoyev2014}.

Within the parametric EOS model outlined in Sect. \ref{eos1}, we generated 40,000 EOSs for neutron stars in the following way: the parameters characterizing the EOS near the saturation density of nuclear matter are fixed at their most probable values known, i.e., $K_{0}=230$ MeV, $E_{\textrm{sym}}(\rho_{0})=31.7$ MeV and $L=58.7$ MeV; while the parameters describing the high-density behavior of nuclear EOS are randomly selected with equal probabilities in the ranges of $-300\leq J_{0} \leq 400$ MeV, $-400\leq K_{\textrm{sym}}\leq 100$ MeV and $-200\leq J_{\textrm{sym}}\leq 800$ MeV, respectively. After removing those (1) violating the causality condition, or (2) can not support the currently observed most massive neutron star of mass $2.01 ~\textrm{M}_{\odot}$\cite{Antoniadis13}, (3) or leading to mechanical instabilities at any density, over 23,000 rational EOSs are left to be used as inputs to solve the TOV and the differential equations governing the complex frequency of the f-mode given in the appendix A.

\begin{figure}
\centering
\includegraphics [width=0.5\textwidth]{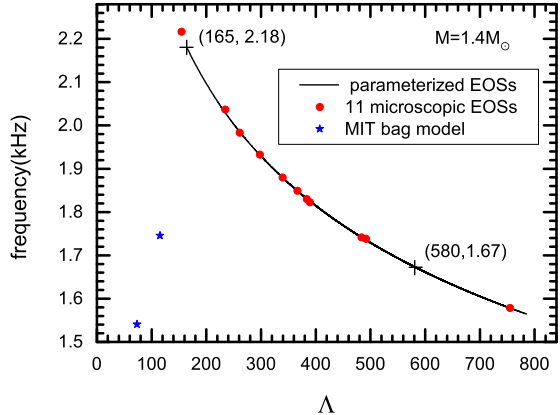}
\caption{ \label{fig1} The correlation between the f-mode frequency and the tidal deformability of canonical neutron stars using 23,000 phenomenological EOSs (solid black line), 11 microscopic EOSs (red dots) and 2 EOSs for quark stars within the MIT bag model (blue stars).  The cross at (165,2.18) corresponds to the lower limit of the tidal deformability predicted using the parameterized EOS satisfying all known constraints from terrestrial nuclear experiments, while the cross at (580, 1.67) corresponds to the upper limit of the tidal deformability of neutron stars extracted from the GW170817 event by LIGO and VIRGO Collaborations.}
\end{figure}

\begin{figure}
\centering
\includegraphics [width=0.5\textwidth]{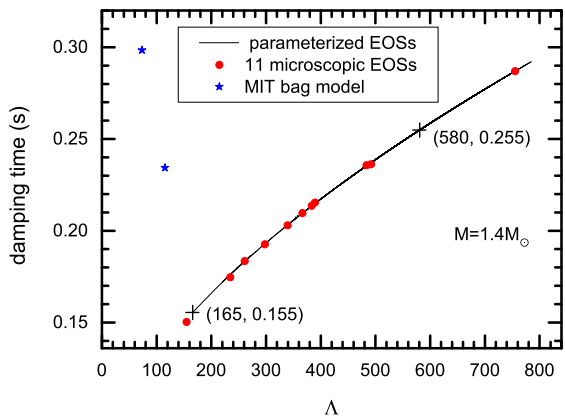}
\caption{ \label{fig2} Same as Fig. \ref{fig1}, but for the f-mode damping time.}
\end{figure}

We now explore possible correlations between the f-mode characteristics and the tidal deformability of neutron stars. Such correlations may exist because it has been well known that both the f-mode frequency and its damping time scale with the compactness M/R independent of the EOS, see, e.g., refs. \cite{Andersson1996,Tsui2005b}, while the tidal deformability has also been found to depend strongly on the $(M/R)^{\alpha}$ although the $\alpha$ is still rather EOS model dependent \cite{Zhang2019}. Our results for a fixed stellar mass of $1.4 ~\textrm{M}_{\odot}$ are presented in Figs. \ref{fig1} and \ref{fig2}. It is very interesting to see that an almost perfect universal correlation exists between the frequency/damping time and the tidal deformability except for the two quark EOSs.  It is not surprising that the $f$ and $\tau$ versus $\Lambda $ relations with the quark star EOSs do not fall onto the same universal relations as with the other EOSs used. Indeed, both the $f$~(or~$\tau)$ versus M/R and $\Lambda$ versus M/R universal relations were suggested earlier without considering any quark star model \cite{Tsui2005b,Yagi2013}. One possible reason is that both the tidal deformability $\Lambda$ and the f-mode characteristics ($f$ and $\tau$) depend strongly on the internal structure of neutron stars. Indeed, the structure of quark stars is quite different from the normal ($npe\mu$ matter with or without hyperons) neutron stars which have a characteristic low-density crust. As a result, their mass-radius relations are very different, leading to the distinct $f$ and $\tau$ versus $\Lambda $ relations.

The universal $f$ and $\tau$ versus $\Lambda $ relations may have some significant applications. Obviously, they enable us to constrain the predicted features of f-mode oscillations using the existing data on the tidal deformability of neutron stars and terrestrial laboratory constrains on the EOS parameters. It is seen that for a fixed stellar mass of $1.4 ~ \textrm{M}_{\odot}$, the lowest tidal deformability $\Lambda$ allowed by the parametric EOS is 165, which is higher than the lower limit of $\Lambda=70$ extracted from analyzing the GW170817 event \cite{LIGO2018}. While the revised maximum tidal deformability $\Lambda=580$ from the improved analyses \cite{LIGO2018} of the GW170817 event limits both the f-mode frequency and damping time.
Therefore, by combining the two constraints: the EOS parameter space allowed by existing terrestrial nuclear laboratory experiments and the tidal deformation extracted from GW170817, the f-mode frequency and damping time for a  $1.4 ~\textrm{M}_{\odot}$ neutron star are constrained in the region of 1.67 kHz - 2.18 kHz and 0.155 s - 0.255 s, respectively, providing a useful guidance for detecting GWs from f-mode oscillations of neutron stars.
As already pointed out by Kokkotas et al. \cite{Kokkotas2001}, in gravitational wave astronomy, the frequency of the f-mode could be detected very precisely, but the error of extracting the damping time from observations could be quite large. As more neutron star mergers are expected to be measured more accurately, the further reduced uncertainty of the tidal deformability especially about its lower limit will constrain more tightly the f-mode features. Moreover, the $f$ and $\tau$ versus $\Lambda $ relations of quark stars deviate significantly from those universal relations of normal neutron stars, providing a possible way of distinguishing the two kinds of neutron stars. Interestingly, this observation is very similar to quark stars' deviations from the universal $I$-$Q$ and $I$-Love number relations \cite{Bandyopadhyay2018}.

To this end, it is very interesting to mention that simulations of merging neutron star binaries have found that the mass scaled frequency (peak frequency) at the maximum amplitude on the spectrum of the post-merger GWs scales approximately universally with some powers of the tidal deformability $\Lambda$ or compactness independent of the EOSs used \cite{Bau12,Rea13,Ber1,Ber2}.
If the remnant formed after the merger can be instantaneously approximated by a perturbed differentially rotating star, the f-mode of pulsation is strongly excited and it is the most efficient channel for GW emissions \cite{Ber2}. Thus, the scaling of the peak frequency in the binary mergers can be related to the f-mode frequency of isolated neutron stars. Indeed, Nils Andersson pointed out that the correlation between the peak frequency and the tidal deformability in neutron star mergers is in principle ``expected" from results of the oscillating single neutron stars \cite{f-custipen}. Clearly, our results presented above support his expectation. The common feature shared by the peak frequency in post-merger spectrum and the f-mode frequency of isolated neutron stars indicate that the differential rotations and thermal effects do not significantly affect the oscillations. Interestingly, it was also found very recently that a strong first-order phase transition will lead to an observable imprint on the gravitational radiation signal from mergers of neutron star binaries \cite{Bauswein2019}. Specifically, the peak frequency was shown to exhibit a significant deviation from an empirical relation between the peak frequency and the tidal deformability if a strong first-order phase transition leads to the formation of a gravitationally stable extended quark matter core in the post merger remnant. 

\section{Scalings of the f-mode frequency and damping time with the compactness of neutron stars of fixed masses}\label{uni}
We first recall here what have been known in the literature about some universal scalings of the f-mode frequency and damping time with respect to the compactness and its variations of neutron stars. We then compare our results from using the 23,000 parametric EOSs with the well established scalings.
To our best knowledge, since the early work of Andersson and Kokkotas \cite{Andersson1996,Andersson1998}, many groups have investigated the universality in characteristics of quasi-normal modes of neutron stars, see, e.g., refs. \cite{Benhar2004,Tsui2005,Wen2009,Lau2010,Blazquez2013,Blazquez2014,Chirenti2015,Stergioulas2018,Brazil}. It is worth noting that all the universal relations reported are not absolutely independent of the EOSs and we focus on the f-mode in this study.

The first widely cited equations to describe the universal relation of f-mode come from the work of Andersson and Kokkotas \cite{Andersson1998}. They found that the f-mode frequency and damping time can be parameterized respectively as
\begin{equation} \label{EQ1}
\omega_{r}\approx \alpha_{r}(\frac{M}{R^{3}})^{1/2}+\beta_{r},
\end{equation}
\begin{equation} \label{EQ2}
 \omega_{i}\approx \frac{M^{3}}{R^{4}}[\alpha_{i}(\frac{M}{R})+\beta_{i}]
\end{equation}
where $\alpha_{r}(=1.635), \beta_{r}(=0.78), \alpha_{i}(=-14.65)$, and $\beta_{i}(=22.85)$ are model independent constants. These universal relations have the features that the frequency $\omega_{r}$ of the  f-mode is proportional to the square root of the stellar average density, while the scaled damping time $M\omega_{i}=M/\tau_{i}$ is a function of the compactness $M/R$. We note that quark stars are not considered in obtaining the above relations.
Two other interesting universal relations were given in refs. \cite{Tsui2005b,Lau2010}. In particular, the following parameterization can be used for most of the quasi-normal modes \cite{Tsui2005b}, including the polar f-mode, 1st polar w-mode, 2nd polar w-mode, 1st axial w-mode and the 2nd axial w-mode. However, it is worth noting that it is almost impossible to observe the w-mode gravitational radiation in actual observations \cite{Allen1998,Allen1999}.
\begin{equation} \label{EQ3}
 M\omega=a(\frac{M}{R})^{2}+b(\frac{M}{R})+c
\end{equation}
where $a, b$, and $c$ are complex constants and have individual values for each mode. For the f-mode, $a=0.15-i5.8\times 10^{-4}$,  $b=0.56+i6.7\times 10^{-4}$, and  $c=-0.020-i6.2\times 10^{-5}$.  It is obvious that the complex eigen-frequency only depends on the compactness $M/R$. This universal relation also does not include quark stars.  The following is their other universal relation given in terms of the effective compactness, $\eta=\sqrt{M^{3}/I}$, where I is the moment of inertia \cite{Lau2010}
\begin{equation} \label{EQ4}
 M\omega_{r}=0.575\eta^{2}+0.133\eta-0.0047,
\end{equation}
\begin{equation} \label{EQ4a}
 M\omega_{i}\eta^{-4}=-0.0256\eta^{2}+0.00694.
\end{equation}
This universal relation works well for both normal neutron stars and quark stars, and have a better accuracy than earlier universal relations. They argue that the effective compactness is a better quantity to characterize the internal mass profile of neutron stars.

Blazquez-Salcedo et al. obtained the following universal relation directly relating the frequency (scaled by the radius) and the damping time (scaled by the mass) \cite{Blazquez2014}
 \begin{eqnarray} \label{EQ5}
&& \omega_{i}/M=-(3.14\pm0.12)\times10^{-10}(\frac{\omega_{r}}{R})^{4}\\
 &+&(164.63\pm0.83)\times10^{-5}(\frac{\omega_{r}}{R})-(5.16\pm0.24)\times10^{-3}.\nonumber
\end{eqnarray}
Again, quark stars are not considered in deriving this relation.

More recently, Stergioulas et al. \cite{Stergioulas2018} extended the universal relation for the damping time of f-mode to higher orders and compared it with that in refs.~\cite{Andersson1998,Tsui2005b}. It reads
\begin{equation} \label{EQ6}
 M\omega_{i}=0.628(\frac{M}{R})^{6}-0.53(\frac{M}{R})^{5}+0.112(\frac{M}{R})^{4}.
\end{equation}
This universal relation has a higher accuracy (with a standard statistical correlation coefficient 0.9997) in a wide range of compactness ($0<M/R<0.33$).

\begin{figure}
\centering
\includegraphics [width=0.5\textwidth]{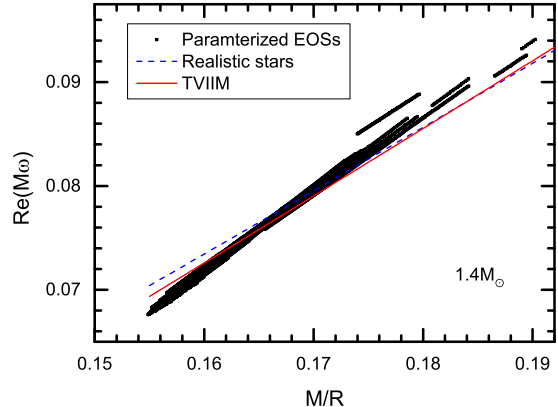}
\caption{ \label{fig3} Scaled f-mode frequency versus compactness of neutron stars with a mass of 1.4 M$_{\odot}$ from using 23,000 parametric EOSs (black points) is compared with the universal relations obtained in reference \cite{Tsui2005b} using the realistic EOSs (blue dashed line) and the Tolman VII model (red line)}
\end{figure}

\begin{figure}
\centering
\includegraphics [width=0.5\textwidth]{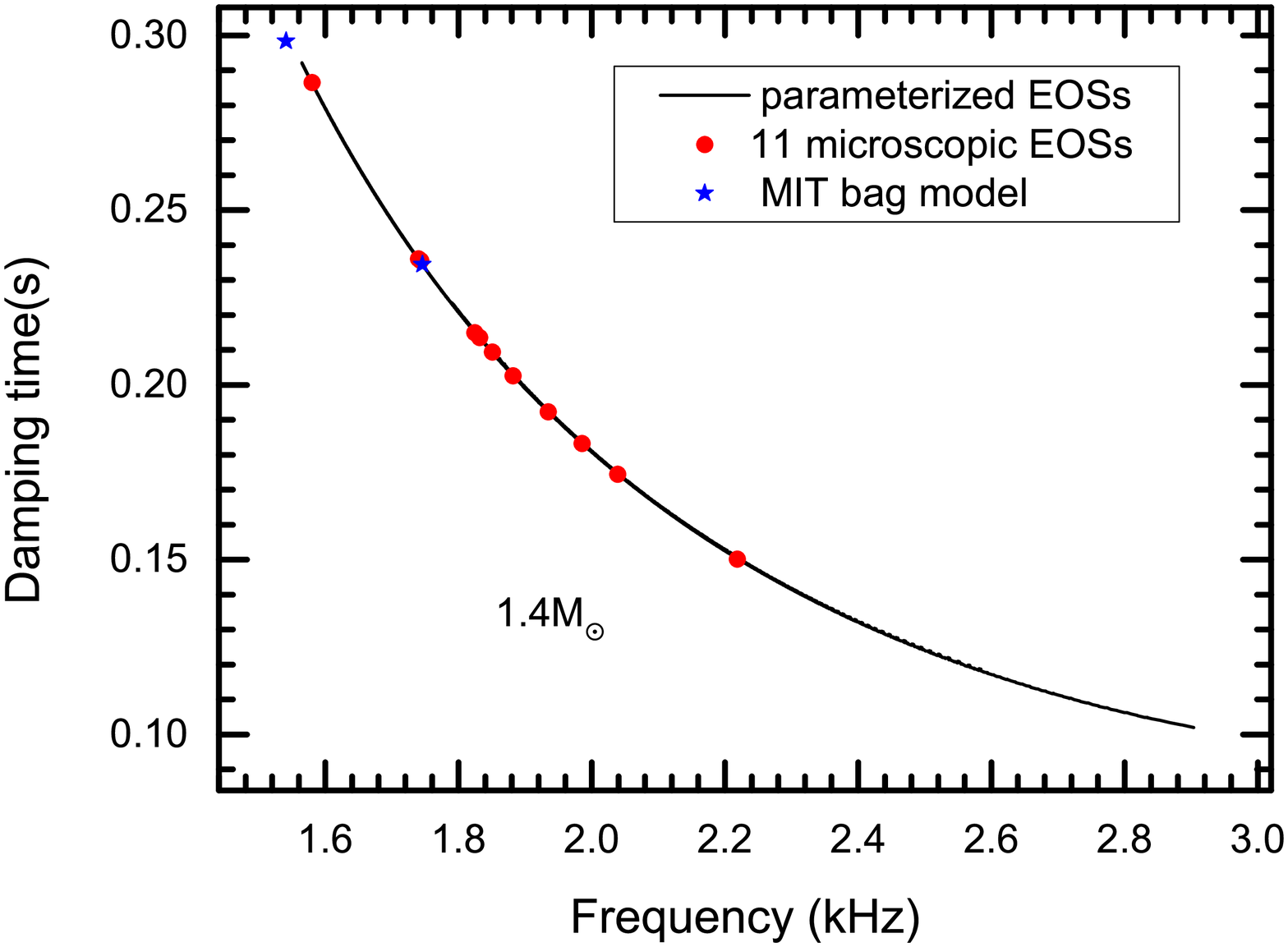}
\caption{ \label{fig4} Correlation between the f-mode frequency and its damping time for neutron stars with a fixed mass of 1.4 M$_{\odot}$ using the 23,000 parametric EOSs, 11 microscopic EOSs and 2 MIT bag model EOSs for quark stars.}
\end{figure}

To this end, it is necessary to compare our results with the well established universal relations mentioned above. Indeed, we found general agreements with the existing relations.
As an example, we compare in Fig. \ref{fig3} our results using the 23,000 parameterized EOSs with the universal relation of Eq. (\ref{EQ3})
given in ref. \cite{Tsui2005b} for a fixed stellar mass of 1.4 M$_{\odot}$. The TVIIM (red line) result was obtained from using a simple TVIIM model (the Tolman VII model \cite{Tolman1939}) to approximate the mass profile inside neutron stars, while the realistic stars (blue dashed line) was obtained by fitting the EOSs predicted by eight different models (APR1,APR2, AU, GM24, MODEL A, MODEL C, UT and UU), for details about these EOS models we refer the reader to refs. \cite{Tsui2005b}. It is seen that our results obtained by using the parameterized EOSs are generally consistent with the universal relation of Eq. (\ref{EQ3}) using the two different sets of EOSs. As we mentioned earlier and shown in Fig.\ \ref{fig3}, the scalings between the frequency and compactness or the average density are not absolutely EOS-independent especially for neutron stars with very low or high compactness.
However, as we shall discuss next and shown in Figs. \ref{fig4}-\ref{fig6}, there is a perfect universal relation between the f-mode frequency and its damping time for neutron stars with fixed masses using all EOSs including
those for quark stars.

\section{Applications of the universal relation between the frequency and damping time of f-mode oscillations of neutron stars with fixed masses}\label{apl}
According to the estimates of Kokkotas et al., the relative error $\sigma_{f}/f$ in extracting the f-mode frequency can be up to $8\times10^{-5}$ assuming a signal-to-noise ratio of 10 \cite{Kokkotas2001}. Thus, the frequency can be measured rather accurately while it is more difficult to extract the damping time. Assuming optimistically both the frequency and/or damping time of the GW emitted from the f-mode oscillation of a neutron star can be accurately measured in the near future, what can we learn about the global properties of neutron stars and the underlying EOS of dense neutron-rich nuclear matter? Here we try to answer this question at least partially to the best we can. In principle, using the parametric EOS one can solve the inverse-structure problem of neutron stars as demonstrated recently in refs. \cite{Zhang2018,Zhang2019,Zhang2019b} to find all necessary combinations of the EOS parameters to produce a given observable or use techniques of Bayesian analyses to infer the probability distribution functions of all EOS parameters from the observational data. As an alternative, here we explore the possibility of inferring global properties of neutron stars and the underlying EOS using the universal relations between the f-mode frequency and its damping time of neutron stars with fixed masses assuming both of them can be obtained accurately from observations.

\begin{figure}
\centering
\includegraphics [width=0.5\textwidth]{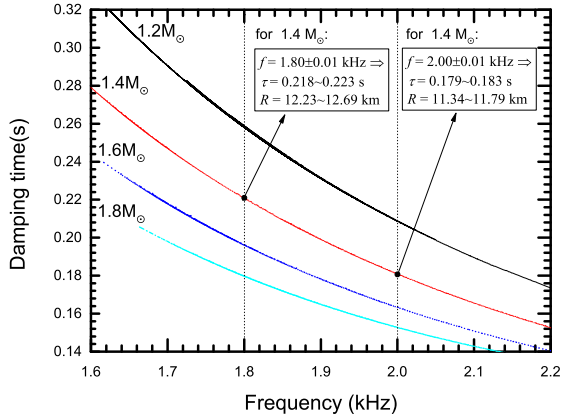}
\caption{ \label{fig5} The universal relations between the f-mode frequencies and damping times for different stellar masses. Constraints on the radii of 1.4 M$_{\odot}$ neutron stars by the supposed frequencies are presented in the two blocks.}
\end{figure}

\begin{figure}
\centering
\includegraphics [width=0.5\textwidth]{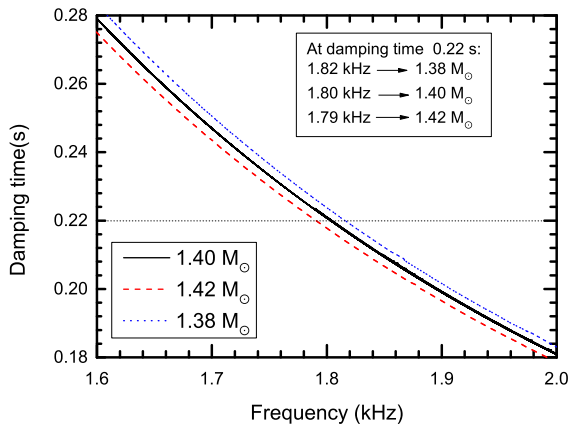}
\caption{ \label{fig6} Resolution for the stellar mass in the  plane of frequency vs damping time by using their universal relations.}
\end{figure}

Using the same sets of EOSs as in Sec. II,  the correlation between the f-mode frequency and its damping time for a canonical neutron star is shown in Fig. \ref{fig4}. It is seen that both the normal neutron stars and quark stars for a given mass fall onto the same universal relation independent of the EOSs used. This finding is not surprising. As outlined in Sect. \ref{fmode}, the differential equations governing the f-mode complex frequency $\omega$ as a whole depend on the EOS through the pressure $p$. Since the real and imaginary parts of the complex frequency $\omega$ vary coherently with the EOS, their dispersion relation for neutron stars with the same mass fall onto the same curve. This is very different from inspecting individually the frequency and damping time as functions of the compactness or tidal deformability which by itself varies with the EOS.
To further illustrate this point, the universal relations with different stellar masses are shown in Figs. \ref{fig5} and \ref{fig6}. As expected, from light to massive neutron stars, the perfect universality between the f-mode frequency and its damping time always holds. Thus, if both the frequency and/or damping time of the GW emitted from the f-mode oscillation of a neutron star can be accurately measured, some global properties of neutron stars can be extracted. For example, if a frequency of $1.80\pm0.01$ kHz is observed from a neutron star of known mass 1.4 M$_{\odot}$,  then the damping time should be 0.218 s - 0.223 s from inspecting the correlations shown in Fig. \ref{fig5}, and the stellar radius should be 12.23 km -12.69 km from inspecting the correlation shown in Figs. \ref{fig3} and \ref{fig5}. Properties corresponding to another example with a measured frequency of $2.00\pm0.01$ kHz are given also in Fig. \ref{fig5}.

Using the results shown in Fig. \ref{fig6} we illustrate what we can learn if both the f-mode frequency and its damping time can be measured but from an isolated neutron star of an unknown mass.
Shown are the universal relations for three neutron stars with similar masses of $M=1.38$ M$_{\odot},~ 1.40$ M$_{\odot} $ and $1.42$ M$_{\odot}$, respectively.
The three relations are clearly separated. If both the frequency $f$ and damping time $\tau$ can be obtained simultaneously with some precisions from future observations, such as, $f=1.80$ kHz and $\tau$=0.22 s, then the stellar mass should be around $1.40$ M$_{\odot}$.

\begin{figure}
\centering
\includegraphics [width=0.5\textwidth]{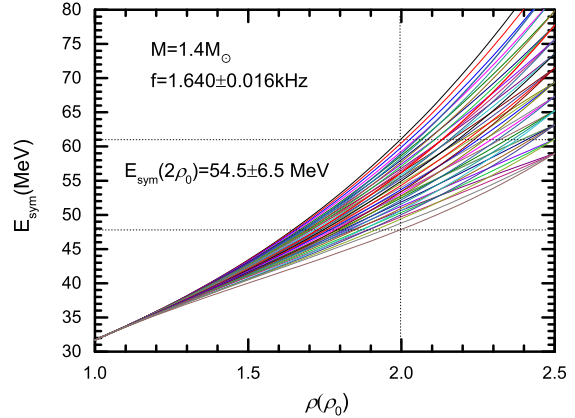}
\caption{ \label{fig7} Constraint on the symmetry energy at 2 times the saturation density of nuclear matter by the supposedly observed f-mode frequency at 1.640 kHz with a 1 percent precision from observing neutron stars of 1.4 M$_{\odot}$.}
\end{figure}

\begin{figure}
\centering
\includegraphics [width=0.5\textwidth]{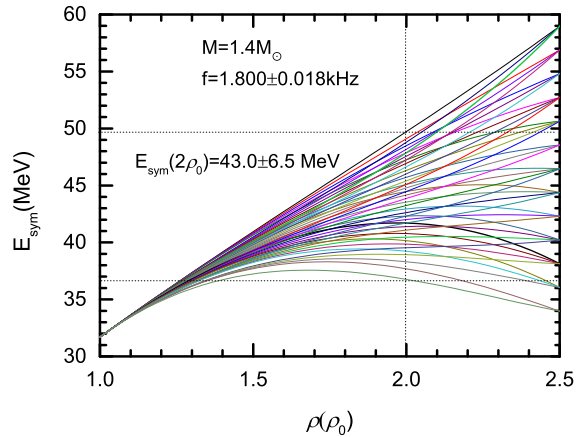}
\caption{ \label{fig8} Same as Fig. \ref{fig7} but with a frequency of 1.800 kHz.}
\end{figure}

One of the major goals of studying properties of neutron stars is to understand and constrain the underlying EOS of dense neutron-rich nuclear matter. While the isospin symmetric part $E_0(\rho)$ of the EOS has been relatively tightly constrained by terrestrial nuclear laboratory experiments \cite{Pawel}, the density dependence of nuclear symmetry energy $E_{\rm sym}(\rho)$ has been the most uncertain part of the EOS of dense neutron-rich nucleonic matter \cite{TEsym}.  While significant progress has been made in probing the $E_{\rm sym}(\rho)$ using nuclear experiments, the high-density behavior of $E_{\rm sym}(\rho)$ remains rather elusive \cite{Li17,TEsym}.
We now explore to what extent the GWs from f-mode oscillations may help constrain the high-density behavior of nuclear symmetry energy.
As examples, we assume that two possible frequencies (1.640 kHz and 1.800 kHz) are detected with a 1\% accuracy for a neutron star of mass 1.4 M$_{\odot}$, then only some values of the high-density symmetry
energy parameters $K_{\rm sym}$ and  $J_{\rm sym}$ are allowed, leading to a constraint on the $E_{\rm sym}(\rho)$ at supra-saturation densities. Our results are shown in Figs. \ref{fig7} and \ref{fig8} for the frequency of 1.640 kHz and 1.800 kHz, respectively. In both cases, the $E_{\rm sym}(\rho)$ spreads further out as the density increases. Since the symmetry energy $E_{\rm sym}(2\rho_0)$ around twice the saturation density $\rho_0$ is especially important for determining the radii of neutron stars \cite{Lattimer01}, it is useful to compare specifically the constraints on the $E_{\rm sym}(2\rho_0)$. If a lower (higher) frequency of 1.640$\pm 0.016$ kHz (1.800$\pm 0.018$ kHz) is observed, then a higher (lower) value of $E_{\rm sym}(2\rho_0)=54.5\pm 6.5$ MeV ($43.0\pm 6.5$ MeV) can be inferred.  Comparing the results in Figs. \ref{fig7} and \ref{fig8}, it is seen that a variation of about 9\% in frequency leads to about 22\% change in the value of $E_{\rm sym}(2\rho_0)$ extracted. Moreover, the tendency of the symmetry energy at even higher densities are also quite different.
To put the numerical results in perspective, we note that an extrapolation of the experimental $E_{\rm sym}(\rho)$ systematics from sub-saturation to supra-saturation densities predicted that $E_{\rm{sym}}({2\rho _{0}}) \approx 40.2 \pm 12.8$ \cite{LWC15}. A recent study of neutron star radii and tidal deformability indicated that $E_{\rm{sym}}(2\rho_0)=46.9\pm10.1$ MeV \cite{Zhang2019b}, while predictions based on nuclear many-body theories scatter between $E_{\rm{sym}}(2\rho_0)=15$ and 100 MeV \cite{Zhang2019b}. Compared to these earlier results especially the rather diverge theoretical predictions, obtaining the limits on the $E_{\rm{sym}}(2\rho_0)$ from the supposed detection of the f-mode frequency with a 1\% accuracy would be a significant progress in constraining the EOS of dense neutron-rich nuclear matter.

\section{Summary}
In summary, within a minimum model for neutron stars using 11 EOSs from microscopic nuclear many-body theories, 2 EOSs from the MIT bag model for quark stars and 40,000  parametric EOSs with their parameters constrained by all existing constraints mostly from terrestrial nuclear laboratory experiments, we studied the correlations among the f-mode frequency and its damping time as well as the tidal deformability and compactness of neutron stars. Besides verifying some of the well established universal relations in the literature, we find that
\begin{itemize}

\item Except for quark stars, both the f-mode frequency and its damping time scale with the tidal deformability of canonical neutron stars independent of the EOSs used.
\item  The tidal deformability $\Lambda_{1.4}=190^{+390}_{-120}$ from analyzing the GW170817 event limits the f-mode frequency and damping time of canonical neutron stars to 1.67 kHz - 2.18 kHz and 0.155 s - 0.255 s, respectively, providing a useful guidance for the ongoing search for gravitational waves from the f-mode oscillations of neutron stars.
\item  The f-mode frequency and its damping time strongly correlate with each other for neutron stars of the same masses.  Assuming optimistically both of them can be obtained with high accuracies from future observations, their correlations allow for the accurate extraction of neutron star global properties and knowledge about the EOS of dense neutron-rich nuclear matter. Several numerical examples under idealized conditions were considered. In particular, nuclear symmetry energy
at twice the saturation density can be extracted with an accuracy compatible with that of several other approaches available.
\end{itemize}

Indeed, gravitational waves from the f-mode oscillations of neutron stars provide useful and complimentary information about both properties of neutron stars and the underlying EOS of dense neutron-rich nuclear matter.

\begin{acknowledgements} We would like to thank Wen-Jie Xie for helpful discussions and Ignacio Francisco Ranea-Sandoval for very useful communications. This work was supported in part by the U.S. Department of Energy, Office of Science, under Award Number DE-SC0013702, the CUSTIPEN (China-U.S. Theory Institute for Physics with Exotic Nuclei) under the US Department of Energy Grant No. DE-SC0009971 and the National Natural Science Foundation of China under Grant Nos. 11275073, 11320101004, 11675226 and 11722546.\\
\end{acknowledgements}

\appendix
\section{The basic formalism for polar f-mode}
According to the work of Lindblom and Detweiler,
the perturbed metric of non-radial oscillating neutron stars
can be written as \cite{Lindblom83,Detweiler85}
\begin{eqnarray}
&ds^{2}=&-e^{\nu}(1+r^{l}H_{0}Y_{m}^{l}e^{i{\omega}t})dt^{2}
-2i{\omega}r^{l+1}H_{1}Y_{m}^{l}e^{i{\omega}t}dtdr\nonumber\\
&&+e^{\lambda}(1-r^{l}H_{0}Y_{m}^{l}e^{i{\omega}t})dr^{2}\nonumber\\
&&+r^{2}(1-r^{l}KY_{m}^{l}e^{i{\omega}t})
(d\theta^{2}+\textrm{sin}^{2}{\theta}d\phi^{2}).
\end{eqnarray}
The fluid element of perturbation can be expressed by the
Lagrangian displacement components as
\begin{equation}
\xi^{r}=r^{l-1}e^{-\lambda/2}WY_{m}^{l}e^{i{\omega}t},\\
\end{equation}
\begin{equation}
\xi^{\theta}=-r^{l-2}V\partial_\theta{Y_{m}^{l}e^{i{\omega}t}},\\
\end{equation}
\begin{equation}
\xi^{\phi}=-r^{l}(r\textrm{sin}\theta)V\partial_\phi{Y_{m}^{l}e^{i{\omega}t}},\\
 \end{equation}
where  the $H_{0}$, $H_{1}$,  $K$,  $W$ and $V$ are perturbation functions,
and the $Y_{m}^{l}$ is the spherical harmonic function. They are not all independent. The perturbed
metric function $H_{0}$ can be represented by other two
perturbed functions as \cite{Lindblom83,Detweiler85}
\begin{widetext}
\begin{eqnarray}
H_{0}&=&[3M+\frac{1}{2}(l+2)(l-1)r+4{\pi}r^{3}p]^{-1} \{8{\pi}r^{3}e^{-\nu/2}-
[\frac{1}{2}l(l+1)(M+4{\pi}r^{3}p)-\omega^{2}r^{3}e^{-(\lambda+\nu)}]H_{1}
\nonumber
\\
& &+[\frac{1}{2}(l+2)(l-1)r-\omega^{2}r^{3}e^{-\nu}
-(M+4{\pi}r^{3}p)(3M-r+4{\pi}r^{3}p)K]\}.
\end{eqnarray}

By using linearized Einstein equations
and the continuity equation, a set of fourth-order differential equations can be deduced as the
following \cite{Lindblom83,Detweiler85}

\begin{eqnarray}
H_{1}^{\prime}&=&-r^{-1}[l+1+2Mr^{-1}e^{\lambda}+4{\pi}r^{2}e^{\lambda}(p-\rho)]H_{1}
                 +r^{-1}e^{\lambda}[H_{0}+K-16\pi(p+\rho)V],\label{9} \\\label{11}
K^{\prime}&=&r^{-1}H_{0}+\frac{1}{2}l(l+1)r^{-1}H_{1}
           -[(l+1)r^{-1}-\frac{1}{2}\nu^{\prime}]K
           -8\pi(p+\rho)e^{\lambda/2}r^{-1}W,\label{10}\\
W^{\prime}&=&-(l+1)r^{-1}W+re^{\lambda/2}[\gamma^{-1}p^{-1}e^{-\nu/2}X
           -(l+1)r^{-1}V+\frac{1}{2}H_{0}],\label{11}\\
X^{\prime}&=&-lr^{-1}X+(p+\rho)e^{\nu/2}\{\frac{1}{2}(r^{-1}-\frac{1}{2}\nu^{\prime})H_{0}
               +\frac{1}{2}[r\omega^{2}e^{-\nu}+\frac{1}{2}l(l+1)r^{-1}]H_{1}\nonumber
                \\
& &+\frac{1}{2}(\frac{3}{2}\nu^{\prime}-r^{-1})K
               -\frac{1}{2}l(l+1)\nu^{\prime}r^{-2}V
               -r^{-1}[4\pi(p+\rho)e^{\lambda/2}+\omega^{2}e^{\lambda/2-\nu}
                \nonumber\\
&
&-\frac{1}{2}r^{2}(r^{-2}e^{-\lambda/2}\nu^{\prime})^{\prime}]W\},\label{12}
\end{eqnarray}
where the prime $'~\prime~'$ represents the first-order derivative with respect to radius,
such as $X^{\prime}=\frac{\displaystyle{dX}}{\displaystyle{dr}}$,
the quantity $\gamma=\frac{\displaystyle{\rho+p}}{\displaystyle{p}}\frac{\displaystyle{dp}}{\displaystyle{d\rho}}$
is the adiabatic index, and the function $X$ (to replace  $V$) is defined as
\begin{equation}
X=\omega^{2}(\rho+p)e^{-\nu/2}V-p^{~\prime}r^{-1}e^{(\nu-\lambda)/2}W
  +\frac{1}{2}(\rho+p)e^{\nu/2}H_{0}.
\end{equation}
To numerically solve the perturbation equations, one can integrate the Eqs. (\ref{9})-(\ref{12}) from the center to the surface of neutron stars. The initial values at the center are taken as
\begin{eqnarray}
W(0)=1, H(0)=K(0)=\pm(\rho_{0}+p_{0}),\\
H_{1}(0)=[2lK(0)+16\pi(\rho_{0}+p_{0})W(0)]/[l(l+1)],\\
X(0)=(\rho_{0}+p_{0})e^{-\nu_{0}/2}\{{[\frac{4\pi}{3}(\rho_{0}+3p_{0})-\omega^{2}e^{-\nu_{0}/2}]W(0)+\frac{1}{2}K(0)}\}.
\end{eqnarray}
\end{widetext}
The unique solution with a specific complex frequency  $\omega  = \omega _{r} + i\omega _{i}$ is determined by the condition $X(R)=0$ at the stellar surface \cite{Kokkotas1999}.
The f-mode oscillation frequency $f=\omega _{r}/(2\pi)$ and its damping time $ \tau=1/\omega _{i}$ are determined by the real and imaginary parts of $\omega$, respectively.
Several methods to numerically solve the above set of  first-order differential equations (Eqs. \ref{11}-\ref{12}) have been proposed, such as the Detweiler-Lindblom method \cite{Lindblom83,Detweiler85}, the Kokkotas-Schutz method \cite{Kokkotas1992} and the Andersson-Kokkotas-Schutz method \cite{Andersson1995}. To  numerically calculate the frequency and damping time of f-mode, we employed the Detweiler-Lindblom method \cite{Lindblom83,Detweiler85} here. The code we used in this work is the same one we used earlier in ref. \cite{Wlin14}.

\end{document}